\documentclass[12pt]{revtex4}
\usepackage{amssymb}
\usepackage{amsmath}
\usepackage{epstopdf}
\usepackage{float}
\usepackage{booktabs}
\usepackage{mathtools}
\usepackage{color}  

\usepackage{physics}
\usepackage{amsbsy}

\newcommand{\half}{\frac{1}{2}}
\newcommand{\negpf}{\frac{\hbar\alpha_0^2}{4}}
\newcommand{\negpfhf}{\frac{\hbar\alpha_0^2}{8}}
\newcommand{\sqhalf}{\frac{1}{\sqrt{2}}}

\newcommand{\tl}[1]{\tilde{#1}}
\newcommand{\tP}{\tl{P}}
\newcommand{\kP}{{\ket{\Psi}}}

\newcommand{\Hel}{{\bf H_{el}}}

\newcommand{\Vso}{{\bf V_{SO}}}
\newcommand{\up}[1]{{a_{{#1}}^{\dag}}}
\newcommand{\down}[1]{{a_{{#1}}}}

\newcommand{\nq}{{x}}

\newcommand{\derv}[1]{#1^{[\nq]}}

\newcommand{\bA}{{\bf A}}
\newcommand{\bAt}{{\bf \tilde{A}}}
\newcommand{\bT}{{\bf \Theta}}
\newcommand{\bLt}{\tilde{\bf L}}

\newcommand{\myHF}{{\mbox{\tiny{HF}}}}
\newcommand{\nG}{\Phi_\myHF}

\newcommand{\EIS}[1]{{\boldsymbol{ \mathbf{#1}}}}
\newcommand{\eis}[2]{{#2_{\sigma  #1}}}
\newcommand{\ups}[1]{{#1_\alpha}}
\newcommand{\dos}[1]{{#1_\beta }}

\newcommand{\sing}[1]{s_{#1}}
\newcommand{\trip}[2]{{t_{#1}^{(#2)}}}
\newcommand{\Xcis}[0]{\tilde{X}}
\newcommand{\Xsoc}[0]{{X}}

\begin{document}
\author{Nicole Bellonzi, Gregory Medders, Evgeny Epifanovsky and Joseph E. Subotnik}
\date{\today} 

\begin{abstract}
For future use in modeling photoexcited dynamics and intersystem crossing, we calculate
spin-adiabatic states and their analytical nuclear gradients within CIS theory.  These energies and forces
should be immediately useful for surface hopping dynamics, which are natural within an adiabatic
framework.  The resulting code has been implemented within the Q-Chem software and preliminary
results suggest that the additional cost of including SOC within the singles-singles block is not
large.
\end{abstract}

\title{Configuration Interaction Singles with Spin-Orbit Coupling: 
Constructing Spin-Adiabatic States and Their Analytical {Nuclear} Gradients}
\maketitle

\section{Introduction}
Intersystem Crossing (ISC) is a key relaxation pathway for many photo-excited systems.
For instance, several aromatic ketones and aldehydes are known to undergo ISC phosphorescence
with almost unity quantum efficiency \cite{itoh:1988:unity}.
As another example, organometallics with heavy elements (e.g. platinum) are also known to undergo
ISC very efficiently\cite{dunietz:2017:jpcc_phosphorescence}. 
Moreover, because of the long lifetimes of triplet states, recent work in photocatalyzed organic
synthesis has sought to isolate organic precursors with propensities to form excited state 
triplets, so as to maximize yields of photocatalyzed product\cite{nicewicz:2016:cr}.

Now, in general, it is well known that triplets tend to be the lowest energy excited states due to 
antisymmetry and exchange\cite{schiffbook}.
Thus, ISC is almost always thermodynamically accessible.
Nevertheless, many molecules do not undergo ISC, highlighting that whether or not a triplet forms after
photoexcitation is dictated {\textbf{ \textit {by dynamics}}} and not by electronic structure. 
Thus, to make predictions about triplet formation, we must run dynamics simulations. 
And, if we are to run nonadiabatic dynamics, the key question to ask is: 
what is the coupling between the singlet and triplet states? 
There are many such operators for only singlet to triplet intersystem crossing, 
including spin-orbit coupling (SOC), hyperfine couplings, and 
spin-spin couplings.  
In the present paper, we will focus on ISC as induced by SOC, for which El-Sayed's rule is 
applicable\cite{elsayed:1963:rule}. 
A few words are now appropriate about the exact form of the SOC operator. 

Formally, the SOC operator is derived  a consequence of the Dirac equation 
and cannot be derived with a non-relativistic theory of quantum
mechanics. 
Nevertheless, up to a factor of two\cite{griffiths}, the form for the  SOC  operator can be rationalized with
straightforward classical
electromagnetic arguments, and nowadays, it is standard within the quantum chemistry community to
use the so-called Breit-Pauli perturbative form of the SOC operator within nonrelativistic quantum
mechanics\cite{abegg:1975:soc}. 
According to the Breit-Pauli form, the SOC is a vector operator with one and two electron components.
The one electron component is:
\begin{eqnarray}
{\Vso} 
&=&
  -\frac{\alpha_0^2}{2} \sum\limits_{j,A}
    \frac{Z_A}{|{\bf r}_{jA}|^3}
    \left({\bf r}_{jA} \cross {\bf p}_j\right)
    \cdot
    {\bf s}_j 
\label{eqn:BreitPauli}
\end{eqnarray}
where $\alpha_0$ is the fine structure constant, and $j$ and $A$ index the electrons and nuclei,
respectively.
$Z_A$ is the charge of nucleus $A$,
${\bf s}_j$ is the spin operator of the $j$ electron, 
${\bf r}_{jA}$ is the distance between electron $j$ and nucleus $A$,
and 
${\bf p}_{j}$ is the momentum of electron $j$.
In this paper, we will restrict ourselves to the one-electron piece of the SOC operator; others have
shown that a screened one electron SOC term can capture many of the effects of the total SOC
operator \cite{marian:2012:wires_isc}.

Now, given an operator for the SOC, suppose we would like to run 
Tully's Fewest Switches Surface Hopping (FSSH)\cite{tully:fssh}
to determine ISC rates and branching ratios. 
This problem has been considered by several authors in recent years 
\cite{martinez:2016:isc_gonzalez, granucci:2012:fssh_spinorbit,thiel:2014:ic_isc}.
For the reader not familiar with Tully's algorithm, a few words about FSSH are now appropriate.
The basic input to an FSSH trajectory are $1$. adiabatic potential energies surfaces, $2$. nuclear
gradients, and $3$. derivative couplings, and the basic ansatz of FSSH is to run dynamics on
adiabatic surfaces, while hopping between surfaces to account for electronic relaxation. 
One key element of surface hopping dynamics is that all dynamics are propagated
along adiabatic surfaces.  
This choice of surface ensures that barrier crossings (without tunneling) are correct and 
also that detailed
balance is preserved approximately\cite{tully:fssh,tully:2005:detailedbalance,tully:2008:detailedbalance}. 
This choice furthermore gives us the correct choice of hopping direction
-- the derivative coupling\cite{herman:1984:direction,arce:1994:direction,kapral:1999:jcp,herman:1995:direction}. 
In fact, the FSSH algorithm can be justified approximately only when the dynamics are run along
an adiabatic basis\cite{subotnik:2013:qcle_fssh_derive}.
{Even when the coupling is small, an adiabatic basis is still feasible
\cite{levine:2014:trivial,shs:1994:protons,tretiak:2013:trivial}.} 
And finally, recent work by several authors has shown that (when studying ISC), the dynamics can
have large errors if one propagates along spin-diabats (i.e. singlet or triplet surfaces) rather
than spin-adiabats (i.e. surfaces that mix spins)\cite{subotnik:2013:qcle_fssh_derive,
kapral:2016:chemphys_fssh, martinez:2016:isc_gonzalez, granucci:2012:fssh_spinorbit}.

Altogether, the evidence above suggests that, in order to run ISC dynamics with FSSH, we should
construct spin-adiabatic wavefunctions, i.e.  the eigenstates of $\Hel + \Vso$ (see Eqn.
\ref{eqn:socA} below)\cite{note:ISC_MS}. 
At this point, however, we must remind ourselves that the cost of diagonalizing the supermatrix will
be large.  
And in fact, we must also recall that mixing one singlet and one triplet does not result in a
system with two electronic states; instead, it results in a system with four electronic states
because the triplets are always triply degenerate; of course all degeneracy is broken by SOC. 
Moreover, it would make no sense to include only one of the three triplets (say $m_s=0$ triplet) because
then the calculation would depend spuriously on the artificial choice of lab frame and molecular
orientation. 
And because the phase of a coupling can be essential (e.g. near a conical intersection),
it is  not reasonable to reduce the system to a single triplet with, e.g. $|\Vso|^2$, as an average
(rotationally invariant)  perturbative matrix element. 
Instead, for a rotationally invariant calculation, one must include all of the components of the
triplet and diagonalize the full Hamiltonian (which also includes all of the vector components of
the SOC operator, see Eqn. \ref{eqn:BPcomp}).  
And yet, given that the most expensive piece of a CIS calculation is multiplication of the trial
amplitudes by the two electron integrals, and given the fact that the two electron integrals do not
mix spin symmetry, it should be possible to compute spin-adiabat electronic states, as well as
spin-adiabatic gradients, with minimal cost{\cite{note:recent_work}}.

With all of this background in mind, our goal for this paper is to derive and implement an algorithm
for quickly generating spin-adiabatic states and their gradients within the context of configuration
interaction singles (CIS), a popular and computationally efficient method to generate excited 
spin-diabatic state energies and amplitudes. 
Similar previous work with a semiempirical approach was performed by Granucci and Persico 
\cite{granucci:2011:jcc}. 

\section{Theoretical Methods}
\subsection{Notation}
Establishing correct notation will be essential for our problem with spatial and spin degrees of
freedom. 
Henceforward,   
lowercase Greek letters $\mu\nu\lambda\gamma\omega$ index atomic orbitals. 
Lowercase Roman letters $pqrs$ index general molecular orbitals from the Hartree Fock ground state 
($\ket{p} = \sum_\mu C_{\mu p} \ket{\mu}$); 
$abcd$ index specifically virtual orbitals,
and $ijklm$ index specifically occupied orbitals.  
Spin orbitals are represented by bold type as $\EIS{p}$ or $\EIS{\mu}$, or when explicitness is
required,
with subscripts as follows:
$\ups{p}  $ for up spin, 
$\dos{p}  $ for down spin, or
$\eis{}{p}$ for either spin.
A single excited state is defined by $\ket{\Phi_\EIS{i}^\EIS{a}} \equiv \up{\EIS{a}}\down{\EIS{i}}\ket{\nG}$. 
The singlet spin-diabat is indexed by ($s$) and triplet spin-diabats ($t$) are indexed by $m_s=-1,0,+1$.
The four spin-diabats (one singlet and three triplets) can be indexed by $\epsilon\in\{s,m_s\}$.
Finally, note that some quantities below will be complex; an asterisk ($*$) will denote the complex
conjugate.

\subsection{Standard CIS}
We begin by outlining standard configuration interaction singles (CIS) theory of excited states. 
{In this work, we use the closed-shell restricted form of the CIS equations.}
The standard CIS algorithm calculates the eigenstates of the electronic
Hamiltonian, 
\begin{eqnarray} 
\Hel  
&=& 
   \sum_{\EIS{pq}} 
     h_{\EIS{pq}} 
     \up{\EIS{p}} \down{\EIS{q}}  
  +\frac{1}{4} \sum_{\EIS{pqrs}} 
     \Pi_{\EIS{pqsr}} 
     \up{\EIS{p}} \up{\EIS{q}} \down{\EIS{r}} \down{\EIS{s}}  
\nonumber\\ 
&=&
   \sum_{\EIS{pq}} 
     \mel**{\EIS{p}}{h}{\EIS{q}} 
     \up{\EIS{p}} \down{\EIS{q}}  
  +\frac{1}{4} \sum_{\EIS{pqrs}} 
     \mel**{\EIS{pq}}{}{\EIS{sr}} 
     \up{\EIS{p}} \up{\EIS{q}} \down{\EIS{r}} \down{\EIS{s}}, 
\end{eqnarray} 
projected into the space of all single excitations:
\begin{eqnarray} 
\tilde{A}_{\EIS{iajb}} 
&=&
   \mel**{\Phi_\EIS{i}^\EIS{a}}{\Hel}{\Phi_\EIS{j}^\EIS{b}} 
\\ 
&=& 
    h_{\EIS{ab}} \delta_{\EIS{ij}} 
  - h_{\EIS{ji}} \delta_{\EIS{ab}} 
  +\delta_{\EIS{ij}} \sum_{\EIS{m}} \Pi_{\EIS{ambm}} 
  -\delta_{\EIS{ab}} \sum_{\EIS{m}} \Pi_{\EIS{jmim}} 
  +\Pi_{\EIS{ajib}} 
  + E_{HF} \delta_{\EIS{ab}} \delta_{\EIS{ij}}.
\label{eqn:cisA1}
\end{eqnarray} 

This quantity can be rewritten in terms of the Fock matrix,
\begin{eqnarray} 
F_{\EIS{pq}} 
&=& 
    h_{\EIS{pq}} 
  +\sum_{\EIS{m}} 
    \Pi_{\EIS{pmqm}} 
\label{eqn:Fock}.
\\
&=& \varepsilon_\EIS{p} \delta_{\EIS{pq}}
\end{eqnarray} 
The energy of orbital $\EIS{p}$ is $\varepsilon_\EIS{p}$.
By inserting Eqn. \ref{eqn:Fock} into Eqn. \ref{eqn:cisA1}, 
we recover the usual CIS theory with ``Hamiltonian'':
\begin{eqnarray} 
\tilde{A}_{\EIS{iajb}}
&=& 
    F_{\EIS{ab}} \delta_{\EIS{ij}} 
  - F_{\EIS{ji}} \delta_{\EIS{ab}} 
  +\Pi_{\EIS{ajib}} 
  + E_{HF} \delta_{\EIS{ab}} \delta_{\EIS{ij}}
\label{eqn:cisA}
\end{eqnarray} 

The CIS amplitudes ${\bf \Xcis}$ solve the following eigenvalue problem, 
\begin{eqnarray} 
\sum_\EIS{bj} 
  \tilde{A}_{\EIS{iajb}} \Xcis_{\EIS{bj}} 
&=& 
  \tilde{E} \Xcis_{\EIS{ai}} 
\end{eqnarray} 
and are normalized such that, 
\begin{eqnarray} 
\sum_\EIS{ai} 
 \Xcis^*_{\EIS{ai}} \Xcis_{\EIS{ai}} 
&=& 
    1
\label{eqn:norm_X} 
\end{eqnarray} 

The CIS Hamiltonian is block diagonal in the basis of spin-diabats (singlets and triplets). 
A CIS singlet state has amplitudes such that
\begin{eqnarray}
\Xcis_{\ups{a}\ups{i}} = \Xcis_{\dos{a}\dos{i}}
\equiv \frac{1}{\sqrt{2}} \sing{ai} 
\end{eqnarray}
A CIS triplet ($m_s=0$) state will have amplitudes
\begin{eqnarray}
\Xcis_{\ups{a}\ups{i}} = - \Xcis_{\dos{a}\dos{i}}
\equiv \frac{1}{\sqrt{2}} \trip{ai}{ 0}
\end{eqnarray}
The remaining CIS triplet states will be degenerate with equivalent amplitudes 
\begin{eqnarray}
\Xcis_{\ups{a}\dos{i}} 
\equiv \trip{ai}{+1}
\\
\Xcis_{\dos{a}\ups{i}} 
\equiv \trip{ai}{-1}
\end{eqnarray}
where
$\trip{ai}{ 0} = \trip{ai}{+1} = \trip{ai}{-1}$. 

\subsection{The Breit-Pauli One-Electron Hamiltonian}
A CIS-SOC Hamiltonian extends CIS by including the SOC through the Breit-Pauli Hamiltonian
${\Vso}$ (Eqn. \ref{eqn:BreitPauli}), which we express here in second quantization notation:
\begin{eqnarray}
{\Vso}_{x} 
&=&
  -\frac{\alpha_0^2}{2} \sum\limits_{pq}
    {\tilde{L_{x}}}_{pq} 
    \cdot \frac{\hbar}{2}
    \left( \up{\ups{p}} \down{\dos{q}} + \up{\dos{p}} \down{\ups{q}} \right) 
\nonumber\\
{\Vso}_{y} 
&=&
  -\frac{\alpha_0^2}{2} \sum\limits_{pq}
    {\tilde{L_{y}}}_{pq} 
    \cdot \frac{\hbar}{2i}
    \left( \up{\ups{p}} \down{\dos{q}} - \up{\dos{p}} \down{\ups{q}} \right) 
\nonumber\\
{\Vso}_{z} 
&=& 
  -\frac{\alpha_0^2}{2} \sum\limits_{pq}
    {\tilde{L_{z}}}_{pq} 
    \cdot \frac{\hbar}{2}
    \left( \up{\ups{p}} \down{\ups{q}} - \up{\dos{p}} \down{\dos{q}} \right) 
\label{eqn:BPcomp}
\end{eqnarray}
Here $\bLt$ captures the angular moment of an electron moving around all of the different 
nuclei $A$ with positions ${\bf r}_A$.
For example,
\begin{eqnarray}
{\tilde{L_{z}}}_{pq}
 =
 \mel**{p}{\bLt_z}{q}
&=& 
   \sum_{A}
    \frac{\hbar Z_A}{i}
    \mel**{p}
     {\frac{[ ({\bf r}-{\bf r}_A) \cross \nabla ]_z}{|{\bf r}-{\bf r}_A|^3}}
     {q} 
\nonumber\\
&=&
   \sum_{A}
    \frac{\hbar Z_A}{i} 
    \left[
     \mel**{p}
      {\frac{ (x-x_A) }{|{\bf r}-{\bf r}_A|^3}\frac{\partial}{\partial y}}
      {q} 
    -\mel**{p}
      {\frac{ (y-y_A) }{|{\bf r}-{\bf r}_A|^3}\frac{\partial}{\partial x}}
      {q} 
    \right]
\end{eqnarray}
One can permute coordinates to recover the $\bLt_x$ and $\bLt_y$ terms.
The $\bLt$ integrals and their derivatives $\derv{\bLt}$ are discussed in the appendix.

Finally, for convenience later on, let us define a compact notation for $\Vso$ in the explicit
spin basis,
\begin{eqnarray}
V_{\eis{}{p}\eis{'}{q}} 
&=& 
  -\frac{\hbar\alpha_0^2}{4} 
    \tilde{L}_{\eis{}{p}\eis{'}{q}} 
\end{eqnarray}
with the following components for $\bLt$, 
\begin{eqnarray}
\tilde{L}_{\ups{p}\ups{q}} 
&=& 
    \tilde{L}_{{z}_{pq}} 
\nonumber\\                                       
\tilde{L}_{\dos{p}\dos{q}} 
&=&
  - \tilde{L}_{{z}_{pq}} 
\nonumber\\                                       
\tilde{L}_{\ups{p}\dos{q}} 
&=&
    \tilde{L}_{{x}_{pq}} 
  +\frac{1}{i}
    \tilde{L}_{{y}_{pq}} 
\nonumber\\                                
\tilde{L}_{\dos{p}\ups{q}} 
&=&
    \tilde{L}_{{x}_{pq}} 
  -\frac{1}{i}
    \tilde{L}_{{y}_{pq}} 
\label{eqn:Ltspin}
\end{eqnarray}



\subsection{The CIS-SOC Hamiltonian}
The CIS-SOC Hamiltonian is the sum of the $\Hel$ and $\Vso$ projected into the space of all single
excitations: 
\begin{eqnarray} 
{A}_{\EIS{iajb}} 
&=& 
   \mel**{\Phi_\EIS{i}^\EIS{a}}
         {\Hel + \Vso}
         {\Phi_\EIS{j}^\EIS{b}}
\nonumber\\
&=&
   \tilde{A}_{\EIS{iajb}}
  +\mel**{\Phi_\EIS{i}^\EIS{a}}
         {\Vso}
         {\Phi_\EIS{j}^\EIS{b}}
\nonumber\\
&=&
    \tilde{A}_{\EIS{iajb}} 
  + V_{\EIS{ab}} \delta_{\EIS{ij}}
  - V_{\EIS{ji}} \delta_{\EIS{ab}} 
\label{eqn:socA}
\end{eqnarray} 
The ${\bAt}$ operator in Eqn \ref{eqn:socA} is the CIS operator (from Eqn. \ref{eqn:cisA}).

Let us construct a CIS-SOC stationary state\cite{note:unbound}, i.e. an eigenstate of ${\bA}$ (not ${\bAt}$)
with energy E: 
\begin{eqnarray} 
\sum_\EIS{bj} 
 {A}_{\EIS{iajb}} 
 \Xsoc_{\EIS{bj}}
&=&
    E
   \Xsoc_{\EIS{ai}} 
\label{eqn:cissoc}
\end{eqnarray} 
The addition of $\Vso$ mixes singlets and triplets. 
Thus, 
such a CIS-SOC eigenstate $\kP$ will have both singlet and triplet contributions,
\begin{eqnarray}
\kP 
&=&
  \sum_{\EIS{ai}}
   \Xsoc_{\EIS{ai}} 
   \ket{\Phi_\EIS{i}^\EIS{a}} 
 = 
  \sum_{{ai}}
  \sum_{\epsilon}
   \Xsoc_{{ai}}^{(\epsilon)}
   \ket{\Phi_{i}^{a(\epsilon)}} 
\nonumber\\
&=& \sqhalf
   \sum_{ai}
    \sing{ai}    
    \left( \ket{\Phi_\ups{i}^\ups{a}} + \ket{\Phi_\dos{i}^\dos{a}} \right) 
  + \sqhalf
   \sum_{ai}
    \trip{ai}{ 0}
    \left( \ket{\Phi_\ups{i}^\ups{a}} - \ket{\Phi_\dos{i}^\dos{a}} \right) 
\nonumber\\
&&+\sum_{ai}
    \trip{ai}{+1} 
    \ket{\Phi_\dos{i}^\ups{a}}
  +\sum_{ai}
    \trip{ai}{-1} 
    \ket{\Phi_\ups{i}^\dos{a}}
\end{eqnarray}
As in standard CIS,
the amplitudes are normalized over all contributions,
\begin{eqnarray}
\sum_\EIS{ai}
 \Xsoc^{*}_{\EIS{ai}}
 \Xsoc_{\EIS{ai}} 
&=&
   \sum_{ai} 
    \sing{ai}^{*} 
    \sing{ai} 
  +\sum_{ai} \sum_{m_s} 
    \trip{ai}{m_s}^{*} 
    \trip{ai}{m_s} 
  =
   1 
\end{eqnarray}

In the explicit spin basis, we can express ${\bf X}$ as,
\begin{eqnarray}
 \Xsoc_{\ups{a}\ups{i}} 
&=&
   \sqhalf
   \left(\sing{ai} + \trip{ai}{ 0}\right) 
\nonumber\\            
 \Xsoc_{\dos{a}\dos{i}} 
&=&
   \sqhalf
   \left(\sing{ai} - \trip{ai}{ 0}\right) 
\nonumber\\            
 \Xsoc_{\ups{a}\dos{i}} 
&=&
   \trip{ai}{+1} 
\nonumber\\            
 \Xsoc_{\dos{a}\ups{i}} 
&=&
   \trip{ai}{-1} 
\label{eqn:Xspin}
\end{eqnarray}
\subsection{Hellmann-Feynman Theory for  the CIS-SOC Gradient}
We can now use Hellmann-Feynman Theory to find an analytical gradient for the CIS-SOC state energy,
$\derv{E}$, given that the CIS-SOC state is an eigenstate of $\bA$:
\begin{eqnarray}
\derv{E} 
&=&
   \sum_{\EIS{abij}}
    \left(
     \Xsoc^{[\nq]*}_\EIS{ai} 
      {A}_{\EIS{iajb}} 
      \Xsoc_{\EIS{bj}}
    +\Xsoc^{*}_{\EIS{ai}} 
      {A}_{\EIS{iajb}} 
      \derv{\Xsoc_{\EIS{bj}}}
    +\Xsoc^{*}_{\EIS{ai}} 
      \derv{{A}}_{\EIS{iajb}} 
      \Xsoc_{\EIS{bj}}
    \right)
\nonumber\\
&=&
  E
  \sum_{\EIS{ai}}
   \left(
    \Xsoc^{[\nq]*}_\EIS{ai} 
     \Xsoc_{\EIS{ai}}
   +\Xsoc^{*}_{\EIS{ai}} 
     \derv{\Xsoc_{\EIS{ai}}}
   \right)
  +\sum_{\EIS{abij}}
    \Xsoc^{*}_{\EIS{ai}}
     \derv{{A}}_{\EIS{iajb}}
     \Xsoc_{\EIS{bj}}
\nonumber\\
&=&
  \sum_{\EIS{abij}}
   \Xsoc^{*}_{\EIS{ai}} 
    \derv{{A}}_{\EIS{iajb}}
    \Xsoc_{\EIS{bj}}
\label{eqn:HFT}
\end{eqnarray}
As in Eqn. \ref{eqn:socA}, the standard CIS electronic Hamiltonian can be separated from the
new SOC terms,
\begin{eqnarray}
\derv{{A}}_{\EIS{iajb}} 
  =
   \derv{\tilde{A}_{\EIS{iajb}}}
  +\derv{V}_{\EIS{ab}} \delta_{\EIS{ij}}
  -\derv{V}_{\EIS{ji}} \delta_{\EIS{ab}}
\label{eqn:Agrad}
\end{eqnarray}
so that,
\begin{eqnarray}
\derv{E} &=&
\sum_{\EIS{abij}}
\Xsoc^{*}_{\EIS{ai}} \derv{\tilde{A}}_{\EIS{iajb}} \Xsoc_{\EIS{bj}}
+\sum_{\EIS{abi}}
\Xsoc^{*}_{\EIS{ai}} \derv{V}_{\EIS{ab}} \Xsoc_{\EIS{bi}}
-\sum_{\EIS{aij}}
\Xsoc^{*}_{\EIS{ai}} \derv{V}_{\EIS{ji}} \Xsoc_{\EIS{aj}}
\label{eqn:dE1}
\end{eqnarray}

So far we have been working in a molecular spin orbital basis, but quantum chemistry algorithms are
usually designed in the atomic spatial orbital basis to take advantage of real-valued Gaussian-type
orbitals with analytic two electron matrix elements. 
To this end, we will now convert to an atomic spin orbital basis, and then later convert to 
an atomic spatial orbital basis.

\subsection{The Atomic Spin Orbital Basis}
 Molecular orbitals are linear
combinations of AOs with coefficients ${\bf C}$. 
Integrals in the molecular orbital basis, such as ${\Vso}$, are sums over integrals calculated in
the atomic orbital basis,
\begin{eqnarray}
V_{\EIS{pq}} 
&=& 
   \sum_\EIS{\mu\nu} 
     C_\EIS{\mu p} 
     V_{\EIS{\mu\nu}} 
     C_\EIS{\nu q}
\label{eqn:Vao}
\end{eqnarray}

For our purposes below, let us define
some important terms in the AO basis,
\begin{eqnarray}
S_\EIS{\mu\nu}
&\equiv&
   \ip{\EIS{\mu}}{\EIS{\nu}} 
\label{eqn:Sspin}
\\
P_\EIS{\mu\nu}
&\equiv& 
   \sum_\EIS{i} 
     C_\EIS{\mu i}
     C_\EIS{\nu i}
\\
\tP_\EIS{\mu\nu} 
&\equiv& 
   \sum_\EIS{p}
     C_\EIS{\mu p} 
     C_\EIS{\nu p}
  =
   P_\EIS{\mu\nu} 
  +\sum_\EIS{a} 
     C_\EIS{\mu a}
     C_\EIS{\nu a} 
\label{eqn:Ptspin}
\\
R_{\EIS{\mu\nu}}
&\equiv&
   \sum_\EIS{ai}
     C_\EIS{\mu a}
     \Xsoc_{\EIS{ai}}
     C_\EIS{\nu i}
\label{eqn:R}
\\
B_{\EIS{\mu\nu}}
&\equiv&
   \sum_\EIS{abi} 
     C_\EIS{\mu a} 
     \Xsoc_{\EIS{ai}}^{*}
     \Xsoc_{\EIS{bi}}
     C_\EIS{\nu b} 
  -\sum_\EIS{aij}
     C_\EIS{\mu j}
     \Xsoc_{\EIS{aj}}
     \Xsoc_{\EIS{ai}}^{*} 
     C_\EIS{\nu i}
\label{eqn:B}
\end{eqnarray}
The first three equations (\ref{eqn:Sspin} - \ref{eqn:Ptspin})
define the overlap matrix, the ground state
density matrix and the formal inverse
${\bf S}^{-1}$, respectively. 
Eqn. \ref{eqn:R} is the CIS amplitudes in the AO basis, also known as the transition density.
Eqn. \ref{eqn:B} is very similar to the difference density matrix, but 
{\textit{the second term is transposed}}. 

In order to convert the derivatives in Eqn. \ref{eqn:dE1} 
from the MO basis to the AO basis,
we will use the $\derv{\Vso}$ term as an
example.
To start, we apply the derivative operator to Eqn. \ref{eqn:Vao},
\begin{eqnarray}
\derv{V}_{\EIS{pq}} 
&=& 
   \sum_{\EIS{\mu\nu}} 
    \derv{V}_{\EIS{\mu\nu}}
     {C}_\EIS{\mu p} 
     {C}_\EIS{\nu q} 
  +\sum_{\EIS{\mu\nu}}
     V_{\EIS{\mu\nu}} 
     \left( 
       \derv{C}_\EIS{\mu p}
       {C}_\EIS{\nu q}
      +{C}_\EIS{\mu p}
       \derv{C}_\EIS{\nu q} 
     \right)
\label{eqn:Vx1}
\end{eqnarray}
The $ \derv{V}_{\EIS{\mu\nu}}$ term is easily dependent on the $\derv{\bLt}$ integrals, which are
directly available, but the $\derv{C}$ terms are not.  
Others have derived the form of the $\derv{C}$ derivatives\cite{maurice:thesis} 
and we summarize the main points here.
The molecular orbital coefficients depend on the overlap of the atomic orbitals 
(${\bf S}$) and the rotation matrix between the virtual and occupied space
($\Theta_{bi}$), so the derivative can we written,
\begin{eqnarray}
\derv{C}_\EIS{\mu p}
&=&
  \sum_\EIS{\gamma \lambda} 
   \frac{\partial C_\EIS{\mu p}}
        {\partial S_\EIS{\gamma\lambda}}
    \derv{S}_\EIS{\gamma\lambda}
  +\sum_\EIS{ck} 
    \frac{\partial C_\EIS{\mu p}}
         {\partial \Theta_\EIS{ck}}
     \derv{\Theta}_\EIS{ck}
\label{eqn:delC1} 
\end{eqnarray}
The form of the partial derivatives can be written
\begin{eqnarray}
\frac{\partial C_\EIS{\mu p}}
     {\partial S_\EIS{\gamma\lambda}}
&=&
  -\half 
    \tP_\EIS{\mu\gamma}
    C_\EIS{\lambda p} 
\label{eqn:dCdS}
\\
\frac{\partial C_\EIS{\mu p}}
     {\partial \Theta_\EIS{ck}}
&=&
    C_\EIS{\mu k} \delta_\EIS{cp}
  - C_\EIS{\mu c} \delta_\EIS{kp}
\label{eqn:dCdT}
\end{eqnarray}
Inserting Eqns. \ref{eqn:dCdS} and \ref{eqn:dCdT} into Eqn. \ref{eqn:delC1}, we find
\begin{eqnarray}
\derv{C}_\EIS{\mu p}
&=&
  -\half \sum_\EIS{\gamma \lambda}
     \tP_\EIS{\mu\gamma}
     \derv{S}_\EIS{\gamma\lambda} 
     C_\EIS{\lambda p} 
  +\sum_\EIS{ck} 
     \left(
       C_\EIS{\mu k} \delta_\EIS{cp} 
      -C_\EIS{\mu c} \delta_\EIS{kp}
     \right)
     \derv{\Theta}_\EIS{ck} 
\label{eqn:delC_fin}
\end{eqnarray}

Finally, inserting Eqn. \ref{eqn:delC_fin} into Eqn. \ref{eqn:Vx1}
gives
\begin{eqnarray}
\derv{V}_{\EIS{pq}} 
&=& 
   \sum_{\EIS{\mu\nu}} 
     \derv{V}_{\EIS{\mu\nu}} 
     {C}_\EIS{\mu p} 
     {C}_\EIS{\nu q} 
\nonumber\\
&&-
   \half \sum_{\EIS{\mu\nu\gamma\lambda}} 
     \derv{S}_\EIS{\gamma\lambda}  
     V_{\EIS{\mu\nu}}
     \left(
       C_\EIS{\lambda p}
       {C}_\EIS{\nu q} 
       \tP_\EIS{\gamma\mu}
      +C_\EIS{\lambda q}
       {C}_\EIS{\mu p}
       \tP_\EIS{\nu\gamma}
     \right)
\nonumber\\
&&+
   \sum_{\EIS{\mu\nu ck}} 
     \derv{\Theta}_\EIS{ck} 
     V_{\EIS{\mu\nu}} 
     \left(
       \left(
         C_\EIS{\mu k} \delta_\EIS{cp} 
        -C_\EIS{\mu c} \delta_\EIS{kp}
       \right) 
       {C}_\EIS{\nu q}
      +{C}_\EIS{\mu p} 
       \left(
         C_\EIS{\nu k} \delta_\EIS{cq} 
        -C_\EIS{\nu c} \delta_\EIS{kq}
       \right)
     \right)
\label{eqn:Vx2}
\end{eqnarray}
Eqn. \ref{eqn:Vx2} is our final form for $\derv{\Vso}$ in the atomic spin orbital basis. 
The same steps can be applied to the standard CIS terms, the Fock term derivatives
 and two electron term derivatives. See reference \cite{fatehi:2011:dercouple}.

\subsection{The CIS-SOC gradient in the Atomic Spin Orbital Basis}
Now we have all the tools required to write the CIS-SOC gradient in the atomic spin orbital basis.
In an atomic spin orbital basis, $\derv{E}$ can naturally be written as the sum of four terms:  
\begin{eqnarray}
\derv{E} 
  =
   \derv{E}_{A}
  +\derv{E}_{V}
  +\sum_{\EIS{ck}} 
     \derv{\Theta}_\EIS{ck} 
     \left(
       Y_\EIS{ck}
      -Y^{SOC}_\EIS{ck}
     \right)
  +\derv{E}_{HF}
\label{eqn:gradspin}
\end{eqnarray}
Let us now define these terms.

The first term $\derv{E}_A$ is the standard CIS contribution without the ground state derivative
$\derv{E}_{HF}$ and without including orbital relaxation\cite{fatehi:2011:dercouple,note:dEA}:
\begin{eqnarray}
\derv{E}_A 
&\equiv&
   \sum_{\EIS{\mu\nu}}
     {B}_\EIS{\mu\nu} 
     \derv{h}_{\EIS{\mu\nu}}  
  +\sum_{\EIS{\mu\nu\lambda\gamma}}
     \derv{\Pi}_{\EIS{\mu\lambda\nu\gamma}}
     \left(
       {B}_\EIS{\mu\nu} 
       P_{\EIS{\lambda\gamma}} 
      +R^*_{\EIS{\mu\nu}} 
       R_{\EIS{\lambda\gamma}}
     \right)
\nonumber\\
&&-
   \half \sum_{\EIS{\mu\nu\gamma\lambda}}
     \derv{S}_\EIS{\gamma\lambda}
     \tP_\EIS{\gamma\mu}
     \left(
       B_\EIS{\lambda\nu}                                                       
      +B_\EIS{\nu\lambda}
     \right)
     F_{\EIS{\nu\mu}}
\nonumber\\
&&-
   \half \sum_\EIS{\mu\nu\lambda\gamma\delta\omega}
     \derv{S}_\EIS{\delta\omega}  
     \tP_\EIS{\lambda\delta} 
     P_\EIS{\omega\gamma}  
     \left(
       {B}_\EIS{\mu\nu} 
      +{B}_\EIS{\nu\mu}
     \right) 
     \Pi_{\EIS{\mu\lambda\nu\gamma}}
\nonumber\\
&&-
   \half \sum_\EIS{\mu\nu\lambda\gamma\delta\omega} 
     \derv{S}_\EIS{\delta\omega}
     \tP_\EIS{\mu \delta} 
     \left(
       R_\EIS{\omega\nu}^* 
       R_\EIS{\gamma\lambda}
      +R_\EIS{\nu\omega}^*
       R_\EIS{\lambda\gamma}
     \right)
     \Pi_{\EIS{\mu\lambda\nu\gamma}}
\nonumber\\
&&-
   \half \sum_\EIS{\mu\nu\lambda\gamma\delta\omega} 
     \derv{S}_\EIS{\delta\omega}
     \tP_\EIS{\gamma \delta} 
     \left(
       R_\EIS{\mu\nu}^*
       R_\EIS{\omega\lambda}
      +R_\EIS{\nu\mu}^*
       R_\EIS{\lambda\omega} \right)
       \Pi_{\EIS{\mu\lambda\nu\gamma}}
\end{eqnarray}

The second term is the contribution from $\Vso$ without including orbital relaxation\cite{note:dEV}:
\begin{eqnarray}
\derv{E}_V 
&\equiv&
  -\negpf \sum_{\EIS{\mu\nu}} 
     \derv{\tilde{L}}_{\EIS{\mu\nu}}
     B_{\EIS{\mu\nu}}
\nonumber\\
&&+
   \negpfhf \sum_{\EIS{\mu\nu\gamma\lambda}}
     \derv{S}_\EIS{\gamma\lambda} 
     \tP_\EIS{\gamma\mu} 
     \left(
       \tilde{L}_{\EIS{\mu\nu}} 
       B_{\EIS{\lambda\nu}} 
      +\tilde{L}_{\EIS{\nu\mu}} 
       B_{\EIS{\nu\lambda}} 
     \right)
\end{eqnarray}

The third term is the gradient component arising from orbital relaxation. Here, again, there are two
terms.
The ${\bf Y}$ term arises from standard CIS theory \cite{fatehi:2011:dercouple}. 
\begin{eqnarray} 
Y_{\EIS{ck}}
&\equiv&
   \sum_\EIS{\mu\nu\lambda\gamma}
     C_\EIS{\lambda c} 
     C_\EIS{\gamma k}
     \left(
       {B}_\EIS{\mu\nu} 
      +{B}_\EIS{\nu\mu}
     \right) 
     \Pi_{\EIS{\mu\lambda\nu\gamma}}
\nonumber\\
&&+
   \sum_\EIS{j \mu\nu\lambda\gamma}
     C_\EIS{\nu j}
     C_\EIS{\mu k}
     \left(
       \Xsoc^*_\EIS{cj} 
       R_\EIS{\gamma\lambda}
      +\Xsoc_\EIS{cj}
       R_\EIS{\gamma\lambda}^*
     \right)
     \Pi_{\EIS{\mu\lambda\nu\gamma}}
\nonumber\\
&&-
   \sum_\EIS{b \mu\nu\lambda\gamma}
     C_\EIS{\nu c} 
     C_\EIS{\mu b} 
     \left(
       R_\EIS{\gamma\lambda}
       \Xsoc^*_\EIS{bk}
      +R_\EIS{\gamma\lambda}^* 
       \Xsoc_\EIS{bk}
     \right)
     \Pi_{\EIS{\mu\lambda\nu\gamma}}
\end{eqnarray} 
The ${\bf Y}^{SOC}$ term arises from the SOC. 
\begin{eqnarray}
Y^{SOC}_\EIS{ck} 
&\equiv&
  -\sum_\EIS{ai}
     \left(
       \Xsoc_{\EIS{c}\EIS{i}}^*
       \Xsoc_{\EIS{a}\EIS{i}} 
       V_{\EIS{k}\EIS{a}}
      +\Xsoc_{\EIS{c}\EIS{i}} 
       \Xsoc_{\EIS{a}\EIS{i}}^*
       V_{\EIS{a}\EIS{k}}
     \right) 
\nonumber\\
&&-
   \sum_\EIS{ai}
     \left(
       V_{\EIS{c}\EIS{i}} 
       \Xsoc_{\EIS{a}\EIS{i}}^* 
       \Xsoc_{\EIS{a}\EIS{k}}  
      +V_{\EIS{i}\EIS{c}} 
       \Xsoc_{\EIS{a}\EIS{i}} 
       \Xsoc_{\EIS{a}\EIS{k}}^*
     \right) 
\label{eqn:Ysocspin}
\end{eqnarray}

The fourth term  is the ground state derivative, $\derv{E}_{HF}$.

\subsection{The Atomic Spatial Orbital Basis}
At this point, we have used only the usual tricks to evaluate the CIS gradient, from Ref. \cite{fatehi:2011:dercouple}.
The last and final step is to integrate over the spin degrees of freedom and
evaluate the gradient in Eqns. \ref{eqn:gradspin}-\ref{eqn:Ysocspin} in terms of atomic spatial orbitals.
This requires explicit spin information in our integrals. 

We emphasize that most of the atomic orbital terms do not mix different spins. The explicit spin
information for these terms can be expressed easily as:
\begin{eqnarray}
C_{\eis{  }{\mu}\eis{ '}{p}}
&=&
   C_{\mu p} \delta_{\sigma \sigma'}
\\
S_{\eis{  }{\mu}\eis{ '}{\nu}}
&=&
   S_{\mu\nu} \delta_{\sigma \sigma'}
\\
P_{\eis{  }{\mu}\eis{ '}{\nu}}
&=& 
   \delta_{\sigma \sigma'}
   \sum_{i} 
     C_{\mu i}
     C_{\nu i}
\\
\tP_{\eis{  }{\mu}\eis{ '}{\nu}}
&=& 
   \delta_{\sigma \sigma'}
   \sum_{p} 
     C_{\mu p}
     C_{\nu p}
\end{eqnarray}
When we integrate over spin, these terms will not contain any spin information. The terms in an atomic
spatial orbital basis that do have unique spin information are,
\begin{eqnarray}
R_{\eis{  }{\mu}\eis{ '}{\nu}}
&=& 
   \sum_{ai}
     C_{\mu a}
     \Xsoc_{\eis{  }{a}\eis{ '}{i}}
     C_{\nu i}
\\
B_{\eis{  }{\mu}\eis{ '}{\nu}}
&=&
   \sum_{abi} \sum_{\sigma''} 
     C_{\mu a} 
     \Xsoc_{\eis{  }{a}\eis{''}{i}}^{*}
     \Xsoc_{\eis{ '}{b}\eis{''}{i}}
     C_{\nu b} 
  -\sum_{aij} \sum_{\sigma''}
     C_{\mu j}
     \Xsoc_{\eis{''}{a}\eis{  }{j}}
     \Xsoc_{\eis{''}{a}\eis{ '}{i}}^{*} 
     C_{\nu i}
\label{eqn:Bspin}
\end{eqnarray}

Now, the ${\bf \Pi}$ tensor might appear more complicated than
necessary
 in the explicit spin basis.
After all, ${\bf \Pi}$ is block diagonal in the spin-diabat basis, and so it 
will be convenient to define the two different forms (depending on spin).
When we integrate over singlet diabatic states, the tensor takes the form,
\begin{eqnarray}
\Pi_{\mu\lambda\nu\gamma}^{(s)}
  \equiv
  2\ip{\mu\lambda}{\nu\gamma} 
  -\ip{\mu\lambda}{\gamma\nu}
\end{eqnarray} 
When we integrate over triplet diabatic states, the tensor takes the form,
\begin{eqnarray}
\Pi_{\mu\lambda\nu\gamma}^{(t)}
  \equiv
  -\ip{\mu\lambda}{\gamma\nu}.
\end{eqnarray}
In this framework, the Fock matrix has a contribution from the singlet form of ${\bf \Pi}$,
i.e.
\begin{eqnarray}
F_{\eis{  }{\mu}\eis{ '}{\nu}}
&=&
   \delta_{\sigma\sigma'}
   \left(
     h_{\mu\nu}
    +\sum_{\lambda\gamma} P_{\lambda\gamma}\Pi_{\mu\lambda\nu\gamma}^{(s)}
   \right)
\end{eqnarray}
For use below, we will also define ${\bf R}$ and ${\bf B}$ in the spin-diabatic basis,
\begin{eqnarray}
R_{\mu\nu}^{(\epsilon)}
&\equiv& 
   \sum_{ai}
     C_{\mu a}
     \Xsoc_{ai}^{(\epsilon)}
     C_{\nu i}
\\
B_{\mu\nu}^{(\epsilon)}
&\equiv&
   \sum_{abi} 
     C_{\mu a} 
     \Xsoc_{ai}^{(\epsilon)*}
     \Xsoc_{bi}^{(\epsilon)}
     C_{\nu b} 
  -\sum_{aij}
     C_{\mu j}
     \Xsoc_{aj}^{(\epsilon)*}
     \Xsoc_{ai}^{(\epsilon)}
     C_{\nu i}
\end{eqnarray}
Here, $\epsilon\in\{s,m_s=-1,0,+1\}$. 

The last matrix element required in a spatial orbital basis  is the 
rotation matrix between the virtual and occupied space. 
Given that we assume the Hartree-Fock ground state will always be a closed shell singlet,
the molecular orbital basis never mixes spin, so that 
\begin{eqnarray}
\Theta_{\eis{  }{c}\eis{ '}{k}}
&=&
   \Theta_{ck} \delta_{\sigma \sigma'}
\label{eqn:thetaspat}
\end{eqnarray}


\subsection{The CIS-SOC gradient in the Atomic Spatial Orbital Basis}
Using the above equations, we can construct a final working analytical gradient:
\begin{eqnarray}
\derv{E} 
  =
   \sum_\epsilon
     \derv{{E_{A}^{(\epsilon)}}}
  +\derv{E}_{V}
  +\sum_{ck} 
     \derv{\Theta}_{ck} 
     \left(
       \sum_\epsilon
         Y^{(\epsilon)}_{ck}
      -\sum_{\sigma}
         Y^{SOC}_{\eis{}{c}\eis{}{k}}
     \right)
  +\derv{E}_{HF}
\label{eqn:gradspat}
\end{eqnarray}
As stated above, the electronic Hamiltonian components are all block diagonal in the 
spin-diabatic basis, and we see the same behavior with the gradient. 
This means we can sum over the independent contributions of the various spin-diabats
($\epsilon\in\{s,m_s=-1,0,+1\}$):
\begin{eqnarray}
\derv{{E_{A}^{(\epsilon)}}}
&\equiv&
   \sum_{\mu\nu}
     {B}_{\mu\nu}^{(\epsilon)} 
     \derv{h}_{\mu\nu}  
  +\sum_{\mu\nu\lambda\gamma}
     \left(
       {\Pi}_{\mu\lambda\nu\gamma}^{(s)[\nq]}
       B_{\mu\nu}^{(\epsilon)}
       P_{\lambda\gamma} 
      +{\Pi}_{\mu\lambda\nu\gamma}^{(\epsilon)[\nq]}
       R_{\mu\nu}^{(\epsilon)*} 
       R_{\lambda\gamma}^{(\epsilon)}
     \right)
\nonumber\\
&&-
   \half \sum_{\mu\nu\gamma\lambda}
     \derv{S}_{\gamma\lambda}
     \tP_{\gamma\mu}
     \left(
       B_{\lambda\nu}^{(\epsilon)}  
      +B_{\nu\lambda}^{(\epsilon)}
     \right)
     F_{\nu\mu}
\nonumber\\
&&-
   \half \sum_{\mu\nu\lambda\gamma\delta\omega}
     \derv{S}_{\delta\omega}  
     \tP_{\lambda\delta} 
     P_{\omega\gamma}  
     \left(
       {B}_{\mu\nu}^{(\epsilon)} 
      +{B}_{\nu\mu}^{(\epsilon)}
     \right) 
     \Pi_{\mu\lambda\nu\gamma}^{(s)}
\nonumber\\
&&-
   \half \sum_{\mu\nu\lambda\gamma\delta\omega} 
     \derv{S}_{\delta\omega}
     \tP_{\mu \delta} 
     \left(
       R^{(\epsilon)*}_{\omega\nu} 
       R^{(\epsilon) }_{\gamma\lambda}
      +R^{(\epsilon)*}_{\nu\omega}
       R^{(\epsilon) }_{\lambda\gamma}
     \right)
     \Pi_{\mu\lambda\nu\gamma}^{(\epsilon)}
\nonumber\\
&&-
   \half \sum_{\mu\nu\lambda\gamma\delta\omega} 
     \derv{S}_{\delta\omega}
     \tP_{\gamma \delta} 
     \left(
       R^{(\epsilon)*}_{\mu\nu}
       R^{(\epsilon) }_{\omega\lambda}
      +R^{(\epsilon)*}_{\nu\mu}
       R^{(\epsilon) }_{\lambda\omega} \right)
       \Pi_{\mu\lambda\nu\gamma}^{(\epsilon)}
\label{eqn:EAxspat}
\end{eqnarray}
Similarly, we can define the ${\bf Y}^{(\epsilon)}$ term,
\begin{eqnarray} 
Y_{ck}^{(\epsilon)}
&\equiv&
   \sum_{\mu\nu\lambda\gamma}
     C_{\lambda c} 
     C_{\gamma k}
     \left(
       {B}_{\mu\nu}^{(\epsilon)} 
      +{B}_{\nu\mu}^{(\epsilon)}
     \right) 
     \Pi_{\mu\lambda\nu\gamma}^{(s)}
\nonumber\\
&&+
   \sum_{j \mu\nu\lambda\gamma}
     C_{\nu j}
     C_{\mu k}
     \left(
       \Xsoc^{(\epsilon)*}_{cj} 
       R^{(\epsilon)}_{\gamma\lambda}
      +\Xsoc^{(\epsilon)}_{cj}
       R^{(\epsilon)*}_{\gamma\lambda}
     \right)
     \Pi_{\mu\lambda\nu\gamma}^{(\epsilon)}
\nonumber\\
&&-
   \sum_{b \mu\nu\lambda\gamma}
     C_{\nu c} 
     C_{\mu b} 
     \left(
       R^{(\epsilon)}_{\gamma\lambda}
       \Xsoc^{(\epsilon)*}_{bk}
      +R_{\gamma\lambda}^{(\epsilon)*} 
       \Xsoc^{(\epsilon)}_{bk}
     \right)
     \Pi_{\mu\lambda\nu\gamma}^{(\epsilon)}
\end{eqnarray} 

For the $\derv{E}_V$ term, we cannot simplify to spin-diabats, so we integrate 
over the spin degrees of freedom and express the results explicitly here,
\begin{eqnarray}
\derv{E}_V 
&=&
  -\negpf \sum_{\mu\nu} \sum_{\sigma\sigma'} 
     \derv{\tilde{L}}_{\eis{  }{\mu}\eis{ '}{\nu}}
     B_{\eis{  }{\mu}\eis{ '}{\nu}}
\nonumber\\
&&+
   \negpfhf \sum_{\mu\nu\gamma\lambda} \sum_{\sigma\sigma'} 
     \derv{S}_{\gamma\lambda} 
     \tP_{\gamma\mu} 
     \left(
       \tilde{L}_{\eis{  }{\mu}\eis{ '}{\nu}} 
       B_{\eis{  }{\lambda}\eis{ '}{\nu}} 
      +\tilde{L}_{\eis{ '}{\nu}\eis{  }{\mu}} 
       B_{\eis{ '}{\nu}\eis{  }{\lambda}} 
     \right)
\label{eqn:spatdEV}
\end{eqnarray}
The $\bLt$ integrals are defined in this basis in Eqn. \ref{eqn:Ltspin}. The ${\bf B}$ terms are defined in 
Eqn. \ref{eqn:Bspin}, using the ${\bf \Xsoc}$ from Eqn. \ref{eqn:Xspin}.
From Eqn. \ref{eqn:thetaspat}, we know that the $\derv{\bT}$ term restricts the spin of ${\bf
Y}^{SOC}$, so we can write this quantity explicitly as,   
\begin{eqnarray}
Y^{SOC}_{\eis{  }{c}\eis{  }{k}} 
&\equiv&
  -\sum_{ai} \sum_{\sigma'\sigma''}
     \left(
       \Xsoc_{\eis{  }{c}\eis{ '}{i}}^*
       \Xsoc_{\eis{''}{a}\eis{ '}{i}} 
       V_{\eis{  }{k}\eis{''}{a}}
      +\Xsoc_{\eis{  }{c}\eis{ '}{i}} 
       \Xsoc_{\eis{''}{a}\eis{ '}{i}}^*
       V_{\eis{''}{a}\eis{  }{k}}
     \right) 
\nonumber\\
&&-
   \sum_{ai} \sum_{\sigma'\sigma''}
     \left(
       V_{\eis{  }{c}\eis{ '}{i}} 
       \Xsoc_{\eis{''}{a}\eis{ '}{i}}^* 
       \Xsoc_{\eis{''}{a}\eis{  }{k}}  
      +V_{\eis{ '}{i}\eis{  }{c}} 
       \Xsoc_{\eis{''}{a}\eis{ '}{i}} 
       \Xsoc_{\eis{''}{a}\eis{  }{k}}^*
     \right) 
\label{eqn:Ysocspat}
\end{eqnarray}

All of the terms in Eqns. \ref{eqn:gradspat}-\ref{eqn:Ysocspat} are 
available in standard quantum chemistry software, except the $\bLt$ integrals
and derivatives, which we have implemented. 
As is standard in 
gradient theory,  
$\derv{\bf \Theta}$ is not calculated directly, 
but rather through the coupled-perturbed Hartree-Fock
theory (CPHF) with a Z-vector scheme\cite{mills:1968:cphf,pople:cphf,handyschaefer:1984:zmatrix}.


\section{Results}
In a development version of the Q-Chem software package\cite{qchem5}, we have 
implemented our CIS-SOC algorithm and the nuclear gradients of these CIS-SOC states (Eqns.
\ref{eqn:gradspat}-\ref{eqn:Ysocspat}). 
Our reference molecule is ethene at the S$_2$/T$_4$ crossing geometry as calculated
at the HF/6-31G** level of theory.

\subsection{States at an Intersystem Crossing}
{
We have made use of a Davidson-inspired iterative diagonalization scheme  for finding stationary
states of the complex CIS-SOC Hamiltonian (Eqn. \ref{eqn:socA}) and satisfy Eqn.
\ref{eqn:cissoc} \cite{davidson:1975,notay:davidson}.}
The method searches for the lowest eigenvalues of a matrix in a subspace of the full basis.
The number of iterations required to converge the eigenvalues depends greatly on the initial guess
of the subspace.
Calculating standard CIS states scales formally as $O(N^2)$ where $N$ is the size of the matrix.
When we mix singlets and triplets, one would naively expect that the computational cost of CIS-SOC
would go up by a factor of 16 relative to standard CIS{; and when one considers the
transition from a real to complex Hermitian Hamiltonian, the cost should go up by another factor of
2 (for a total factor of 32 times the cost)}. Perhaps {not} surprisingly,
we have found that the cost of CIS-SOC is reduced dramatically if we use
standard CIS singlets and triplets as an initial guess subspace.

For our small example of ethene, all
calculations were run in serial. When one uses 5 singlet and 5 triplet standard CIS states 
as the initial guess (20 spin-diabats in total), the CIS-SOC calculation requires only 
4 iterations. Table \ref{table:cpu} shows that the
total calculation time for this example is less than the cost of standard CIS.

\begin{table}[H]
\centering
\begin{tabular}{c|ccc}
   & Wall Time & CPU Time & Iterations\\\hline
CIS      & 5.6 s & 3.3 s & 29\\
CIS-SOC  & 3.8 s & 3.5 s & 4\\
Total    & 9.4 s & 6.8 s & - 
\end{tabular}
\caption{Calculation timing at the S$_2$/T$_4$ crossing geometry of ethene.
For this case, the CIS-SOC calculation does not add much overall cost to a CIS calculation.}
\label{table:cpu}
\end{table}



\begin{figure}[H]
  \centering
    \includegraphics[width=0.5\textwidth]{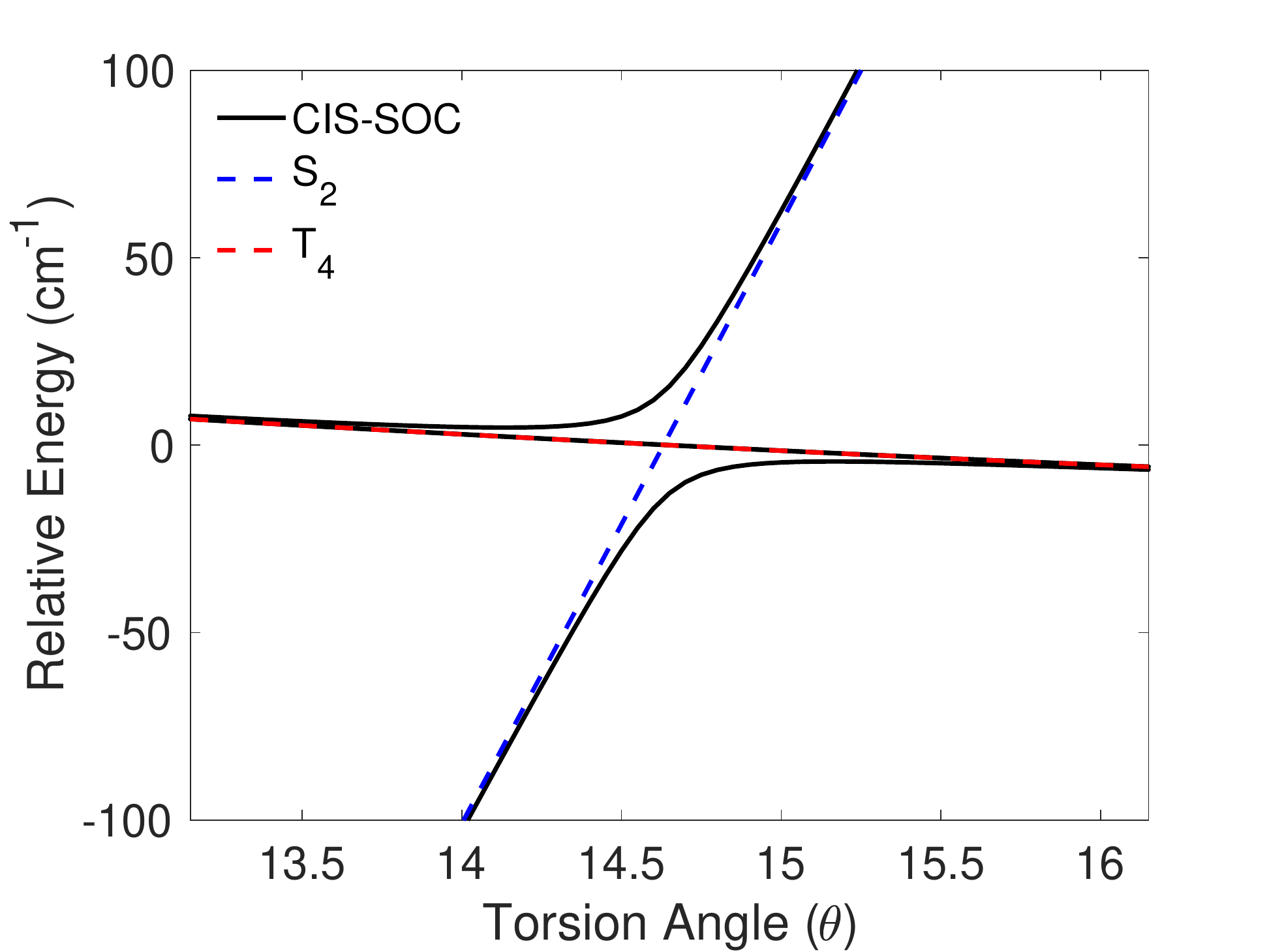}
    \caption{ \label{fig:1} 
The CIS and CIS-SOC potential energy surfaces at the S$_2$/T$_4$ intersystem crossing of ethene as a
function of the torsion angle 
relative to the energy of the crossing point of the CIS states ($\tilde{E}(\theta = 14.635) = 77.676$ Ha). 
The blue (red) dashed line represents the singlet (triplet) CIS spin-diabat.
The black lines are the CIS-SOC spin-adiabats (states $11-14$). 
The crossing point geometry is given in the appendix.
}  
\end{figure}

Figure \ref{fig:1} demonstrates the effect of including the SOC at the S$_2$/T$_4$ crossing geometry. 
When one rotates ethene along its torsion
angle, the CIS energies of the S$_2$ and T$_4$ 
states cross at $\theta_c \equiv 14.635^{\circ}$ and $\tilde{E}(\theta_c) = 77.675$ E$_h$. 
With SOC, the S$_2$ singlet diabat and the three T$_4$ triplet diabats mix, 
generating two mixed spin-adiabats and two degenerate triplet
adiabats. 

\begin{figure}[H]
  \centering
    \includegraphics[width=0.25\textwidth]{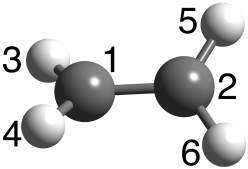}
    \caption{ \label{fig:geo} 
A schematic representation of ethene twisting along its torsion angle. The numbers correspond to 
the labels used in the appendix.
}  
\end{figure}

\subsection{Comparison to Finite Difference}
For the ethene case in Fig. \ref{fig:1}, we have used a {five point} stencil to calculate energy gradients
by finite difference at the crossing point for H3 and C1. 

\begin{table}[H]
\centering
\begin{tabular}{l|lll|lll}
     & \multicolumn{3}{c}{Finite Difference} & \multicolumn{3}{c}{Analytical}\\
     & \multicolumn{3}{c}{ (E$_h\cdot a_0^{-1}$)} & \multicolumn{3}{c}{ (E$_h\cdot a_0^{-1}$)}\\
Atom &    x       &   y       &   z       &    x         &   y       &    z       \\ \hline
H3   &  -0.00530  &  0.01884  &  0.00761  &    -0.00530  &  0.01884  &  0.00761   \\
C1   &  -0.12767  &  0.00000  &  0.00000  &    -0.12767  &  0.00000  &  0.00000   
\end{tabular}
\caption{Gradient of CIS-SOC state  14 in atomic units. Note that the analytical results agree with finite difference.}
\label{table:FD}
\end{table}

\section{Discussion and Conclusion}
We have derived and implemented analytic gradients for the spin-adiabatic states corresponding to a
CIS Hamiltonian when we include SOC. As argued in the introduction, there are many applications for
which we believe this theory will be relevant, especially surface hopping 
nonadiabatic dynamics. Nevertheless, the approach taken here has been by brute force, and one might
wonder if the math to get to Eqns. \ref{eqn:gradspat}-\ref{eqn:Ysocspat} was really necessary? 
After all, if we want spin-adiabats, one must wonder why we have not implemented 
the most obvious alternative algorithm?
Naively, we could calculate the singlet and triplet states directly and then
couple a smaller subset together through SOC
\cite{gonzalez:2014:pert,krylov:2015:soc_cc,thiel:2017:soc_lrt,granucci:2012:fssh_spinorbit,granucci:2011:jcc}.
With such an approach,
however, we emphasize that
one cannot apply Hellmann-Feynman theorem,
so that for a 
derivative,
one must calculate explicitly
how the singlet and triplet states change
as a function of nuclear coordinates\cite{granucci:2011:jcc}. 
Furthermore, using a Z-vector to address such changes may be unstable due to 
high energy intruder states or \cite{alguire:2013:closs_for_yarkony}. 
By contrast, our presented method with spin-adiabats avoids all such difficulties; while we spend
somewhat more time on matrix diagonalization, we spend far less time on the gradient. 

Looking forward, one big question is how to transfer all this technology from CIS to TD-DFT.
After all, TD-DFT is known to
correct the orbital energies relative to Hartree-Fock and CIS, and yield much better
excitation energies. Of course, there are problems with charge transfer states, but using TD-DFT, many
problems can be resolved if one uses a range corrected functional
\cite{savin:1995:book, savin:1995:ijqc_lrc, gill:1996:molphys_lrc, hirao:2004:jcp_lrc, hirao:2007:jcp_lrc, handy:2004:cpl_lrc, scuseria:2006:jcp_lrc, scuseria:2008:jcp_lrc, chaimhg:2008:jcp_lrc, herbert:2008:jcp_lrc, herbert:2009:jcp_lrc, herbert:2008:jpcb_lrc, baer:2010:arpc}. 
Now, when calculating spin-adiabats with TD-DFT, the most obvious difficulty is how to treat the 
exchange-correlated functional which looks like a two-electron matrix element 
in the singles-singles block. However, for such an operator to be nonzero
all electrons must have the same spin. Thus, formally, one should  recover
different excitation energies for the $m_s = 1$ or $-1$ triplets relative to the $m_s = 0$ triplet
(and the latter is more accurate).
Nevertheless,one solution to this quandary would be to simply include the same exchange-correlation
functional for all triplet terms, which will necessarily maintain the normal spin degeneracy and
should produce better excitation energies.

Finally, in the future, one can now imagine several applications worth exploring.
With a fast enough ab initio code,
an obvious target is the photophysics of benzaldehyde and benzophenone and the resulting ISC and 
phosphorescence \cite{ou:2013:benzaldehyde,lindh:2016:jpcl_benzophenone}.
More generally, there have recently been interesting experiments done by Vinogradov and coworkers, 
where two singlets converted to triplets in platinum complexes and there have been few calculations
\cite{vinogradov:2013:tt_annihilation}. 
These are just two out of many possible future applications.

\section{Acknowledgement}
We thank 
Yihan Shao
for helpful conversations. This material is based upon
work supported by the
the National Science Foundation under Grant No. CHE-1764365 
and the National Science Foundation
Graduate Research Fellowship Program under Grant No. DGE-1321851.

\section{Appendix}
\subsection{Geometry of the S$_2$/T$_4$ Crossing Point of Ethene}
The S$_2$/T$_4$ crossing point was found by optimizing the geometry of the ground state
and rotating two geminal hydrogens. Table \ref{table:geo} reports the geometry at the crossing point. 

\begin{table}[H]
\centering
\begin{tabular}{c|ccc}
 Atom  & x ($\AA$) & y ($\AA$)&  z ($\AA$)\\\hline
C (1)   &      0.658180  &   0.000000  &   0.000000\\
C (2)   &     -0.658180  &   0.000000  &  -0.000000\\
H (3)   &      1.224535  &  -0.907854  &  -0.116580\\
H (4)   &      1.224537  &   0.907854  &   0.116580\\
H (5)   &     -1.224537  &  -0.907854  &   0.116580\\
H (6)   &     -1.224537  &   0.907854  &  -0.116580
\end{tabular}
\caption{Geometry of ethene at the S$_2$/T$_4$ Crossing Point}
\label{table:geo}
\end{table}

\subsection{The Spin-Orbit Coupling Integrals and Their Derivatives}
The CIS-SOC gradient requires access to the {derivatives of the spin-orbit integrals} 
in the atomic orbital basis,
$\derv{\tilde{\bf L}}$ in Eqn. \ref{eqn:spatdEV}. 
To this end, we have implemented King and Furlani's algorithm \cite{KF88} for {spin-orbit integrals}
and extended {the algorithm to integral nuclear derivatives.}
King and Furlani present {recursion relations} that express the 
{spin-orbit} multidimensional
integrals as products of one dimensional integrals summed over the roots of the Rys polynomial
\cite{KF88}. 
Here, we will quickly outline the 
relevant {formulas} to evaluate the
nuclear derivatives of the {integrals}.  

\subsubsection{Spin-orbit coupling in terms of nuclear attraction integrals}
Atomic orbitals are linear combinations of Guassian primitive functions of the form, 
\begin{eqnarray}
\ket{\eta_a} = (x-x_a)^{n^x_a}(y-y_a)^{n^y_a}(z-z_a)^{n^z_a} e^{- \alpha_a {\bf r}_a^2}
\end{eqnarray}
Here ${\bf{r}}_a = |r-r_a|$ is the distance from the center of the Gaussian 
and the exponents $\{n^x, n^y, n^z\}$ are the angular momentum for each coordinate and
are elements of the natural numbers $\mathbb{N}_0$. 
Consider
the contribution of an atom $C$ to the $\tilde{L}_z$ term of the SOC 
(of course, one can permute coordinates to recover the $\tilde{L}_x$ and $\tilde{L}_y$ terms).
\begin{eqnarray}
\mel**{\eta_a}{\tilde{L}_{z,C}}{\eta_b} = 
\frac{\hbar Z_C}{i} \left[
 \int{\eta_a}{\frac{(x-x_C)}{{\bf{r_C}}^3}\frac{\partial {\eta_b}}{\partial y}}dr
-\int{\eta_a}{\frac{(y-y_C)}{{\bf{r_C}}^3}\frac{\partial {\eta_b}}{\partial x}}dr
\right]
\end{eqnarray}
King and Furlani convert the integrals with $1/{\bf r}^3$ dependence to 
nuclear attraction integrals ($1/{\bf r}$ dependence) of the form,
\begin{eqnarray}
\mel**{\eta_a}{L_{z,C}}{\eta_b} = 
\frac{\hbar Z_C}{i} \left[
 \mel**{\eta_a^{[x]}}{\frac{1}{{\bf{r_C}}}}{\eta_b^{[y]}}   
-\mel**{\eta_a^{[y]}}{\frac{1}{{\bf{r_C}}}}{\eta_b^{[x]}}   
\right]
\end{eqnarray}
where the derivative of the Gaussian can be defined as
\begin{eqnarray}
\ket{\eta_a^{[x]}} \equiv \frac{\partial {\eta_a}}{\partial x}
=
 n^x_a \ket{\eta_a^{x-}} - 2 \alpha_a \ket{\eta_a^{x+}}
\label{eqn:derivative}
\end{eqnarray}
using short hand for increasing or decreasing the angular momentum in a coordinate $x$:
\begin{eqnarray}
\ket{\eta_a^{x\pm}} 
\equiv
(x-x_a)^{n^x_a\pm1}(y-y_a)^{n^y_a}(z-z_a)^{n^z_a} e^{-\alpha_a {\bf{r}}_a^2}
\end{eqnarray}

\subsubsection{Extension to the Nuclear Derivative}
The nuclear derivative for the $q$ coordinate of atom $D$ of  $\tilde{\bf L}_{z,C}$ is 
\begin{eqnarray}
\mel**{\eta_a}{L_{z,C}}{\eta_b}^{[q_D]}
= \mel**{\eta_a}{L_{z,C}}{\eta_b^{[q_D]}}
+ \mel**{\eta_a}{L_{z,C}^{[q_D]}}{\eta_b}
+ \mel**{\eta_a^{[q_D]}}{L_{z,C}}{\eta_b}
\label{eqn:ap_Luvq}
\end{eqnarray}
The first (last) term only contributes if $\eta_{b}$ ($\eta_{a}$) is centered on atom $D$. The
middle term only contributes if $C\equiv D$. If all three contribute, the integral is zero by
translational invariance.

If $\eta_a$ is centered on atom $D$, we use the properties of Gaussians (Eqn. \ref{eqn:derivative})
to write
\begin{eqnarray}
 \mel**{\eta_a^{[q_D]}}{L_{z,C}}{\eta_b} &=&  
-\mel**{\eta_a^{[q]}}{L_{z,C}}{\eta_b}   
\nonumber\\&=&
 - n^q_a \mel**{\eta_a^{ q-}}{L_{z,C}}{\eta_b}   
 + 2 \alpha_a \mel**{\eta_a^{ q+}}{L_{z,C}}{\eta_b}   
\label{eqn:ap_nuq}
\end{eqnarray}
The same can be said of $\eta_b$.

To evaluate the $L_{z,D}^{[q_D]}$ term, we can use the translational invariance of the integral:
\begin{eqnarray}
\mel**{\eta_a}{L_{z,D}^{[q_D]}}{\eta_b}
&=&
- \mel**{\eta_a^{[q_A]}}{L_{z,D}}{\eta_b}   
- \mel**{\eta_a}{L_{z,D}}{\eta_b^{[q_B]}}   
\nonumber\\
&=&
  \mel**{\eta_a^{[q]}}{L_{z,D}}{\eta_b}   
+ \mel**{\eta_a}{L_{z,D}}{\eta_b^{[q]}}   
\nonumber\\
&=&
   n^q_a \mel**{\eta_a^{ q-}}{L_{z,C}}{\eta_b}   
 - 2 \alpha_a \mel**{\eta_a^{ q+}}{L_{z,C}}{\eta_b}   
\nonumber\\
&&+n^q_a \mel**{\eta_a}{L_{z,C}}{\eta_b^{ q-}}   
 - 2 \alpha_a \mel**{\eta_a}{L_{z,C}}{\eta_b^{ q+}}   
\label{eqn:ap_Lq}
\end{eqnarray}
Thus, if one can evaluate the integrals with different values of angular momentum, one can also
easily evaluate the gradient. 

Let us consider two examples. 
The first example is when $C=D$ and $\eta_a$ and $\eta_b$ are not centered on atom $C$.
\begin{eqnarray}
\mel**{\eta_a}{L_{z,C}}{\eta_b}^{[q_C]}
&=&
\mel**{\eta_a}{L_{z,C}^{[q_C]}}{\eta_b}
\nonumber\\
&=&
   \mel**{\eta_a^{[q]}}{L_{z,C}}{\eta_b}
  +\mel**{\eta_a}{L_{z,C}}{\eta_b^{[q]}}
\end{eqnarray}
The second example is when $C=D$ and $\eta_a$ is centered on $C$, but $\eta_b$ is not.
\begin{eqnarray}
\mel**{\eta_c}{L_{z,C}}{\eta_b}^{[q_C]}
&=&
   \mel**{\eta_c^{[q_C]}}{L_{z,C}}{\eta_b}
  +\mel**{\eta_c}{L_{z,C}^{[q_C]}}{\eta_b}
\nonumber\\
&=&
  -\mel**{\eta_c^{[q]}}{L_{z,C}}{\eta_b}
  +\left(\mel**{\eta_c^{[q]}}{L_{z,C}}{\eta_b}
  +      \mel**{\eta_c}{L_{z,C}}{\eta_b^{[q]}}\right)
\nonumber\\
&=&
   \mel**{\eta_a}{L_{z,C}}{\eta_b^{[q]}}
\end{eqnarray}

\bibliographystyle{apsrev-nourl}
\bibliography{finalbib,notes}{}

\begin{thebibliography}{65}
\bibitem[{not({\natexlab{c}})}]{note:unbound}
\bibinfo{note}{We emphasize that we are not including the SOC operator when we
  optimize orbitals for the original HF calculation; as is well known, such an
  approach would be unbounded\cite{marion::unbounded}. That being said, no such
  problem should arise in practice if SOC is included only within a post-HF CIS
  calculation.}
\bibitem[{not({\natexlab{d}})}]{note:dEA}
\bibinfo{note}{Formally, $\derv{E}_{A}$ is defined $\derv{E}_A \equiv
  \sum_{\EIS{abij}} \Xsoc^{*}_{\EIS{ai}} \derv{{A}}_{\EIS{iajb}}
  \Xsoc_{\EIS{bj}} -\sum_{\EIS{ck}} \derv{\Theta}_\EIS{ck} Y_\EIS{ck}
  -\derv{E}_{HF}$.}
\bibitem[{not({\natexlab{e}})}]{note:dEV}
\bibinfo{note}{Formally, $\derv{E}_{A}$ is defined $ \derv{E}_V \equiv
  \sum_{\EIS{abi}} \Xsoc^{*}_{\EIS{ai}} \derv{V}_{\EIS{ab}} \Xsoc_{\EIS{bi}}
  -\sum_{\EIS{aij}} \Xsoc^{*}_{\EIS{ai}} \derv{V}_{\EIS{ji}} \Xsoc_{\EIS{aj}}
  +\sum_{\EIS{ck}} \derv{\Theta}_\EIS{ck} Y^{SOC}_\EIS{ck} $.}
\end{thebibliography}
\end{document}